\documentclass[twocolumn,preprintnumbers,superscriptaddress,amsmath,amssymb,prb]{revtex4-2}

\usepackage{graphicx}
\usepackage{dcolumn}
\usepackage{subfigure}
\usepackage{bm}
\usepackage{inputenc}  
\usepackage[unicode=true, bookmarks=false, breaklinks=false, pdfborder={0 0 1},colorlinks=false]{hyperref}
\usepackage{breakurl}
\usepackage{url}
\usepackage{natbib}
\usepackage{multirow}
\usepackage{float}

\usepackage{xcolor}
\usepackage{soul}

\begin{document}

\preprint{S.Ghorai et al./V-FMPS}

\title{Site-specific atomic substitution in a giant magnetocaloric $Fe_2P$-type system}

\newcommand{\uppsala}{Department of Materials Science and Engineering,Uppsala University, Box 35, SE-751 03,Uppsala, Sweden}
\newcommand{\schem}{Department of Materials and Environmental Chemistry, Stockholm University, SE-10691 Stockholm, Sweden}
\newcommand{\phys}{Department of Physics and Astronomy, Uppsala University, Box 516, SE-751 20 Uppsala, Sweden}
\newcommand{\chem}{Department of Chemistry – Ångstr{\"o}m Laboratory, Uppsala University, Box 538, Uppsala, 75121, Sweden}
\newcommand{\kth}{Department of Materials Science and Engineering, Royal Institute of Technology, Stockholm SE-100 44, Sweden}
\newcommand{\china}{Faculty of Materials Science and Chemistry, China University of Geosciences, Wuhan 430074, China
}

\author{Sagar Ghorai}
 \affiliation {\uppsala}
\email[]{sagar.ghorai@angstrom.uu.se}

\author{Johan Cedervall}
\affiliation{\schem}

\author{Rebecca Clulow}
\affiliation{\chem}

\author{Shuo Huang}
\affiliation{\china}
\affiliation{\kth}

\author{Tore Ericsson}
\affiliation{\chem}

\author{Lennart Häggström}
\affiliation{\chem}

\author{Vitalii Shtender}
\affiliation{\chem}

\author{Erna K. Delczeg-Czirjak}
\affiliation{\phys}

\author{Levente Vitos}
\affiliation{\kth}

\author{Olle Eriksson}
\affiliation{\phys}

\author{Martin Sahlberg}
\affiliation{\chem}

\author{Peter Svedlindh}
\affiliation{\uppsala}

\date{\today}

\begin{abstract}
Giant magnetocaloric (GMC) materials constitute a requirement for near room temperature magnetic refrigeration. (Fe,Mn)$_2$(P,Si) is a GMC compound with strong magnetoelastic coupling. The main hindrance towards application of this material is a comparably large temperature hysteresis, which can be reduced by metal site substitution with a nonmagnetic element. However, the (Fe,Mn)$_2$(P,Si) compound has two equally populated metal sites, the tetrahedrally coordinated $3f$ and the pyramidally coordinated $3g$ sites. The magnetic and magnetocaloric properties of such compounds are highly sensitive to the site specific occupancy of the magnetic atoms. Here we have attempted to study separately the effect of $3f$ and $3g$ site substitution with equal amounts of vanadium. Using formation energy calculations, the site preference of vanadium and its influence on the magnetic phase formation are described. A large difference in the isothermal entropy change (as high as 44\%) with substitution in the $3f$ and $3g$ sites is observed. The role of the lattice parameter change with temperature and the strength of the magnetoelastic coupling on the magnetic properties are highlighted.
\end{abstract}

\maketitle


\section{Introduction}

Replacement of conventional vapour compression refrigeration  with a $20-30\%$ more efficient solid state magnetic refrigeration technique based on the magnetocaloric effect  has the additional advantage of reducing emission of  greenhouse gases \cite{gschneidner2008thirty}. To build a magnetic refrigerator which can work near room temperature, materials with a giant magnetocaloric (GMC) effect and a magnetic phase transition temperature near room temperature are required. In this regard, several GMC materials with first order magnetic phase transitions have been proposed \cite{Lyubina_2017}. Despite of high values of the isothermal entropy change and adiabatic temperature change, these first order materials may not be suitable for magnetic refrigeration  owing to a large temperature hysteresis ($\Delta T_{hys}$). $\Delta T_{hys}$ represents the irreversible nature of the temperature dependent magnetic phase change, which is a drawback for magnetic refrigeration \cite{gutfleisch2016mastering}. While several ways of reducing $\Delta T_{hys}$ have been attempted   \cite{gutfleisch2016mastering,cohen2018contributions}, the basic origin of $\Delta T_{hys}$ is still unclear. Here in this work we provide an explanation for the origin of $\Delta T_{hys}$ in the context of magnetoelastic coupling of (Fe,Mn)$_2$(P,Si)-type materials.
(Fe,Mn)$_2$(P,Si)-type materials constitute a class of GMC materials consisting of earth abundant, environment friendly and non-toxic elements. These compounds crystallize in a hexagonal Fe$_2$P-type structure (space group $P\overline62m$). In the hexagonal structure the metallic atoms occupy the $3f$ and $3g$ sites while the non metallic atoms occupy the $1b$ and $2c$ sites \cite{miao2014tuning,miao2016short}. From electronic structure calculations \cite{dung2011mixed}, the observed magnetoelastic coupling for this series of compounds has been explained by a drastic fall of the magnetic moment of Fe ($1.54  \mu_{B}/$atom to $\sim 0.003  \mu_{B}/$atom) while it transforms from the ferromagnetic (FM) to the paramagnetic (PM) state. This moment change occurs due to the fact that the non-bonded or metallic Fe below the Curie temperature ($T_C$)  hybridizes with Si/P above $T_C$. The hybridization around $T_C$ causes a drastic change of the hexagonal lattice parameters and a strong magnetoelastic coupling results \cite{dung2011mixed}. For GMC materials, a high magnetization is beneficial, which is mainly provided by the Mn atoms. It can be stated as that the Fe atoms maintain the first order phase transition, while Mn atoms maintain the overall magnetization of (Fe,Mn)$_2$(P,Si)-type materials. To understand the effect of these two phenomena, Fe and Mn atoms are individually attempted to be replaced with non-magnetic V in this work. Recently Lai et al.  \cite{lai2019combined,lai2020tuning} have discussed the reduction of $\Delta T_{hys}$ with V substitution in the metallic sites. However, the occupation of V in the metallic sites (i.e. $3f$ or $3g$ sites) is still unclear. Interestingly, the magnetic atoms exhibit  different magnetic moments depending upon their site occupancy \cite{hudl2011strongly}, yielding completely different magnetic and magnetocaloric properties (e.g. $T_C$, saturation magnetization ($M_S$), isothermal entropy change ($-\Delta S_{M}$), $\Delta T_{hys}$ etc.). Here, in this work we have attempted to substitute the $3g$ and $3f$ sites of the parent compound FeMnP$_{0.5}$Si$_{0.5}$ individually with 5 at$\%$ of V. Both site substitutions exhibit small difference in the $T_C$ values, while there is a large difference ($\sim 44\%$ at $\mu_{0}H = 2$T) in the value of $-\Delta S_{M}$. This difference is explained by the strength of the magnetoelastic coupling and the preferred occupancy of V.              
 
\section{Experimental details and calculation method}
  All compounds were synthesized by the drop synthesis method \cite{hoglin2015phase}. Further, the vacuum sealed samples (pressed pellets) were sintered at $1373$ K for $1$ hr, followed by annealing at $1073$ K for $65$ hrs before quenched in ice water. X-ray powder diffraction (XRPD) data were collected at different temperatures ranging from $265$ K to $422$ K using a Bruker D$8$ Advance diffractometer with Cu-K$_{\alpha1}$ radiation,  with an angle step size of $0.02^{\circ}$. Variable temperature XRD data were analysed using Pawley refinements within the topas 6 software program \cite{topas2018topas}. M{\"o}ssbauer measurements were carried out on a constant acceleration spectrometer with a  $^{57}$CoRh source. The samples were enclosed in sealed kapton pockets yielding a sample concentration of $\approx10$ mg/cm$^{2}$. Calibration spectra were recorded at $295$ K using natural Fe metal foil as a reference absorber. The spectra were recorded at $410$ K and fitted using the least square M{\"o}ssbauer fitting program Recoil to obtain the values of the center shift \textit{CS}, the magnitude of the electric quadrupole splitting $\mid$\textit{QS}$\mid$, the full-width at half maxima \textit{W} of the Lorentzian absorption lines and the spectral intensities \textit{I}. 
  Magnetic properties were measured in the temperature range from $5$K to $400$K using Quantum Design MPMS-XL and PPMS systems with a maximum magnetic field of $5$T. EDX (energy dispersive X-ray) measurements were performed on a Zeiss Leo $1550$ field emission SEM (scanning electron microscope) equipped with Aztec energy dispersive X-ray detector. Data were collected on at least $10$ spots of each sample using an accelerating voltage of $20$ kV by and EDx mapping was carried out on regions of approximately $300~\mu m \times 300~\mu m$. 
  
  The total energy calculations were carried out by the exact muffin-tin orbitals (EMTO) method in combination with the coherent potential approximation (CPA) \cite{vitos2007computational}. The one-electron Kohn-Sham equation was solved within the soft-core and scalar-relativistic approximations. The $s$, $p$, $d$ and $f$ orbitals were included in the muffin-tin basis set. The Green's function was calculated by using $16$ complex energy points on a semicircular contour including the valence states. The exchange-correlation interactions were treated within the generalized gradient approximation in the form of Perdew-Burke-Ernzerhof (PBE) \cite{perdew1996generalized}. Further details about the adopted method can be found in previous work \cite{vitos2007computational}. For the further discussion the parent compound, FeMnP$_{0.5}$Si$_{0.5}$ and the two V substituted compounds FeMn$_{0.95}$V$_{0.05}$P$_{0.5}$Si$_{0.5}$ ($3$g site substituted) and Fe$_{0.95}$V$_{0.05}$MnP$_{0.5}$Si$_{0.5}$ ($3f$ site substituted) are abbreviated as P, V$3g$, and V$3f$, respectively.

\section{Results and discussion}

\subsection{Magnetoelastic coupling and magnetocaloric effect}
The magnetocaloric effect is often characterized with the isothermal entropy change ($-\Delta S_{M}$). The total entropy of a system is the sum of magnetic, lattice and electronic entropy contributions of the system. For a first order magnetic phase transition (i.e. a system with a discontinuity in the first order derivative of the Gibbs free energy) the  magnetic phase transition is often associated with a lattice or electronic phase transition. Moreover, if this is the case, a high value of $-\Delta S_{M}$ is expected.

\begin{figure*}[ht]
    \centering
    \includegraphics[width=\linewidth]{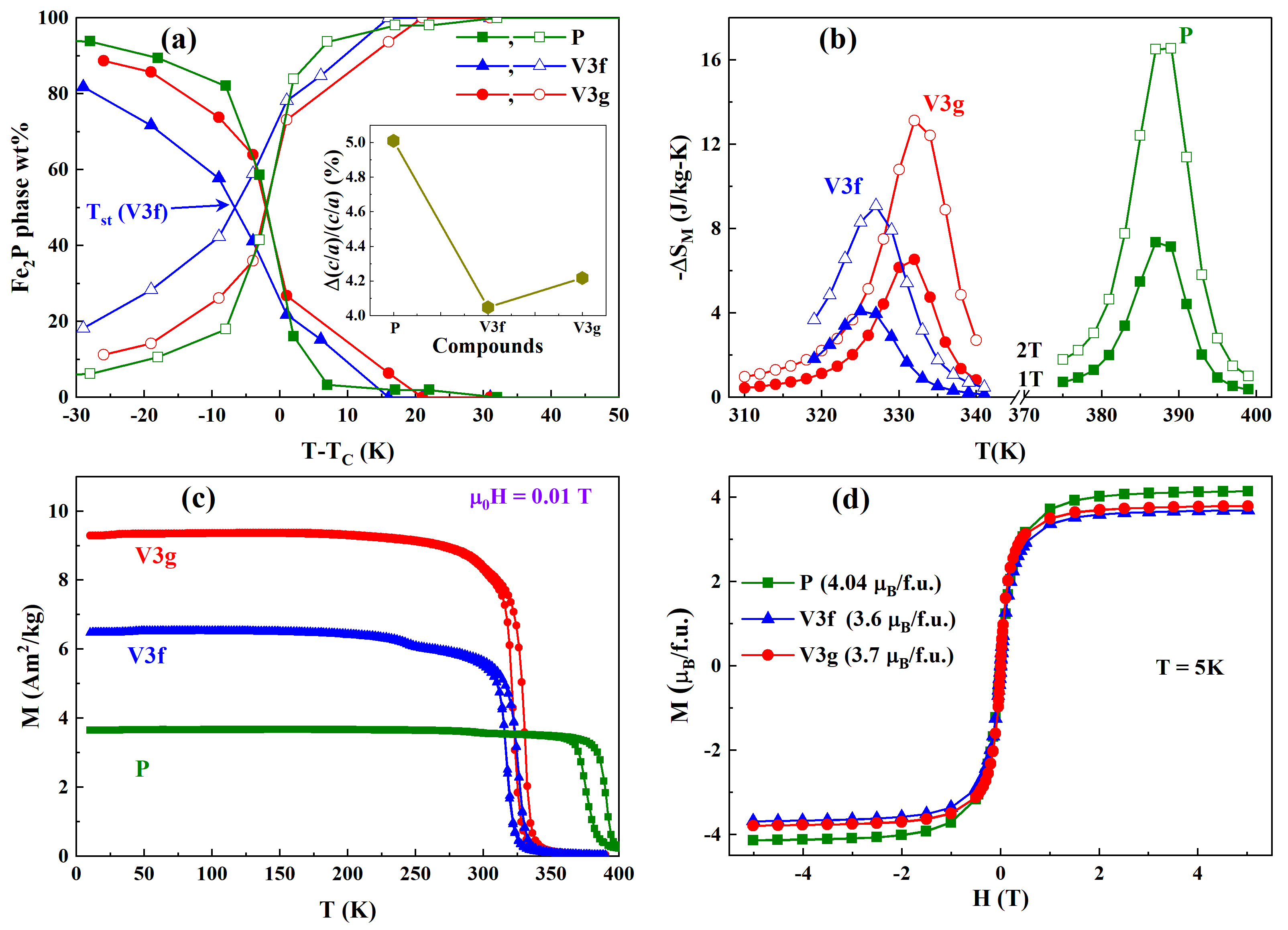}
    \caption{(a) Temperature dependent variation of Fe$_2$P phases with low ($\sim 0.53$) and high ($\sim 0.57$) $c/a$ ratios represented by solid and hollow symbols,  respectively. The crossing points of the two Fe$_2$P phases represent structural transition temperature $T_{st}$ (for V$3f$ compound, the blue arrow indicates the $T_{st}$) of the studied compounds. The inset shows the relative variation of $c/a$ for the different compounds transforming from the PM to the FM state. (b) Temperature dependent variation of the isothermal entropy change, hollow (solid) symbols represent data at a magnetic field change of $2$ T ($1$ T). (c) Temperature dependent variation of magnetization at $\mu_{0}H = 0.01$ T. (d) Isothermal magnetization at $5$ K.} 
    \label{MCE} 
\end{figure*}

The (Fe,Mn)$_2$(P,Si)-system shows a first order magnetoelastic phase transition, where a sharp change of the hexagonal lattice parameter ratio $c/a$ (keeping the lattice volume almost constant) occurs in the vicinity of the magnetic phase transition. Hence, the total entropy change of the system includes by both magnetic and lattice entropy contributions.  In our studied compounds a temperature dependent change of the $c/a$ ratio from a high value ($\sim 0.57$) to a relatively lower value ($\sim 0.53$) has been observed when the system transforms from the PM to the FM state. At high temperature ($>T_{C}+30$ K) the system corresponds fully to a Fe$_2$P-type phase with a high $c/a$ ratio, while at a sufficiently low temperature ($<T_{C}-100$ K) the system corresponds fully to a Fe$_2$P-type phase with a low $c/a$ ratio. Therefore, near $T_{C}$, contributions (phase wt$\%$) from both Fe$_2$P-type phases with high and low $c/a$ ratios  are present. From the collected XRPD patterns, the temperature dependent variation of two Fe$_2$P-type phases are shown in Fig.\ref{MCE}(a).The crossing point of the two Fe$_2$P phases represent the $c/a$ transition temperature, i.e. the structural phase transition temperature ($T_{st}$, cf. Fig.\ref{MCE}(a)). The strength of the magnetoelastic coupling depends upon two factors; firstly the degree of structural change, in this case the variation of the $c/a$ ratio, which is shown in the inset of Fig.\ref{MCE}(a), and secondly the difference between the structural and magnetic phase transition temperatures. From the relative difference between $T_{st}$ and $T_C$ (cf. Fig.\ref{MCE}(a)) and the change of absolute value of $c/a$ (cf. inset of Fig.\ref{MCE}(a)), it is clear that the magnetoelastic coupling strength is highest for parent compound (P) followed in descending order by the V$3g$ and V$3f$ compounds. As discussed before, the strength of magnetoelastic coupling has a direct influence on the value of $-\Delta S_{M}$. The value of $-\Delta S_{M}$ has been calculated using the Maxwell relation \cite{gschneidner2005recent}, \(\Delta S_{M}(T,H_f) = \mu_{0}\int_{0}^{H_f}(\frac{\delta M (H,T)}{\delta T})_{H} \,dH \), for a magnetic field change of $H_{f}$. The magnetic isotherms have been recorded during cooling of the material with a cyclic measurement protocol \cite{caron2009determination}, where the sample is subsequently heated to its PM state before stabilizing at the temperature of the measurement. The calculated values of $-\Delta S_{M}$ are presented in Fig.\ref{MCE}(b). The values of $-\Delta S_{M}$ follows the same trend as that of the magnetoelastic coupling strength, i.e. the heighest value for parent compound followed by lower values of the V$3g$ and V$3f$ compounds. This is a manifestation of the  proportionality between the magnetoelastic coupling strength and $-\Delta S_{M}$.

To estimate the temperature range of the materials to be useful as a magnetic refrigerant, the relative cooling power ($RCP$) is often used \cite{ghorai2020field}. The $RCP$ value can be calculated from the temperature dependent $-\Delta S_{M}$ curve as \(RCP= -\Delta S_{M}^{max} \times \Delta T_{FWHM}\), where $-\Delta S_{M}^{max}$ is the maximum of the isothermal entropy change and $\Delta T_{FWHM}$ is the full width at half maximum of the temperature dependent $-\Delta S_{M}$ curve. Here the $RCP$ and $-\Delta S_{M}$ values of the studied compounds along with the corresponding data for some well known GMC materials are listed in Table\ref{MCEtable}.

\begin{table}[ht]
\caption{Magnetocaloric properties of the studied compounds (*) compared with data reported for other GMC materials near room temperature.}
\vspace{3pt}
\centering

\begin{ruledtabular}
\begin{tabular}{cccccc}
 Sample&   $T_{C}^{FC}$&   $\mu_{0}H$&  $-\Delta S_{M}$ & $RCP$ & Ref.\\
&(K)&(T) & (J/kgK)& (J/kg)& \\
\hline
FeMnP$_{0.5}$Si$_{0.5}$(P)&      376&    2&    16.5&   147&    *\\
Fe$_{0.95}$V$_{0.05}$MnP$_{0.5}$Si$_{0.5}$(V3f)&    318&    2&   9.1&   103&    *\\
FeMn$_{0.95}$V$_{0.05}$P$_{0.5}$Si$_{0.5}$(V3g)&    322&    2&   13.1&   130&    *\\
Gd&      295&    2&    6.1&   240&    \cite{gschneidner2000magnetocaloric}\\
La(Fe$_{0.98}$Mn$_{0.02}$)$_{11.7}$Si$_{1.3}$H&      312&    2&    13&   -&    \cite{shen2009recent}\\
La$_{0.67}$Ca$_{0.33}$MnO$_3$&      260&    1.5&    4.3&   47&    \cite{phan2007review}\\
La$_{0.5}$Pr$_{0.2}$Ca$_{0.1}$Sr$_{0.2}$MnO$_3$&      296&    2&    1.8&   147&    \cite{skini2020large}\\
Fe$_{80}$Pt$_{20}$&      290&    2&    $\sim 10$&   -&    \cite{rong2007temperature}\\
Mn$_{1.2}$Fe$_{0.8}$P$_{0.75}$Ge$_{0.25}$&      288&    2&    20&   -&    \cite{trung2009tunable}\\
MnFeP$_{0.52}$Si$_{0.48}$&      268&    2&    10&   -&    \cite{cam2008structure}\\

\end{tabular}
\end{ruledtabular}

\label{MCEtable}
\end{table}

\subsection{Curie temperature and hysteresis}

\begin{figure}[ht]
    \centering
    \includegraphics[width=\linewidth]{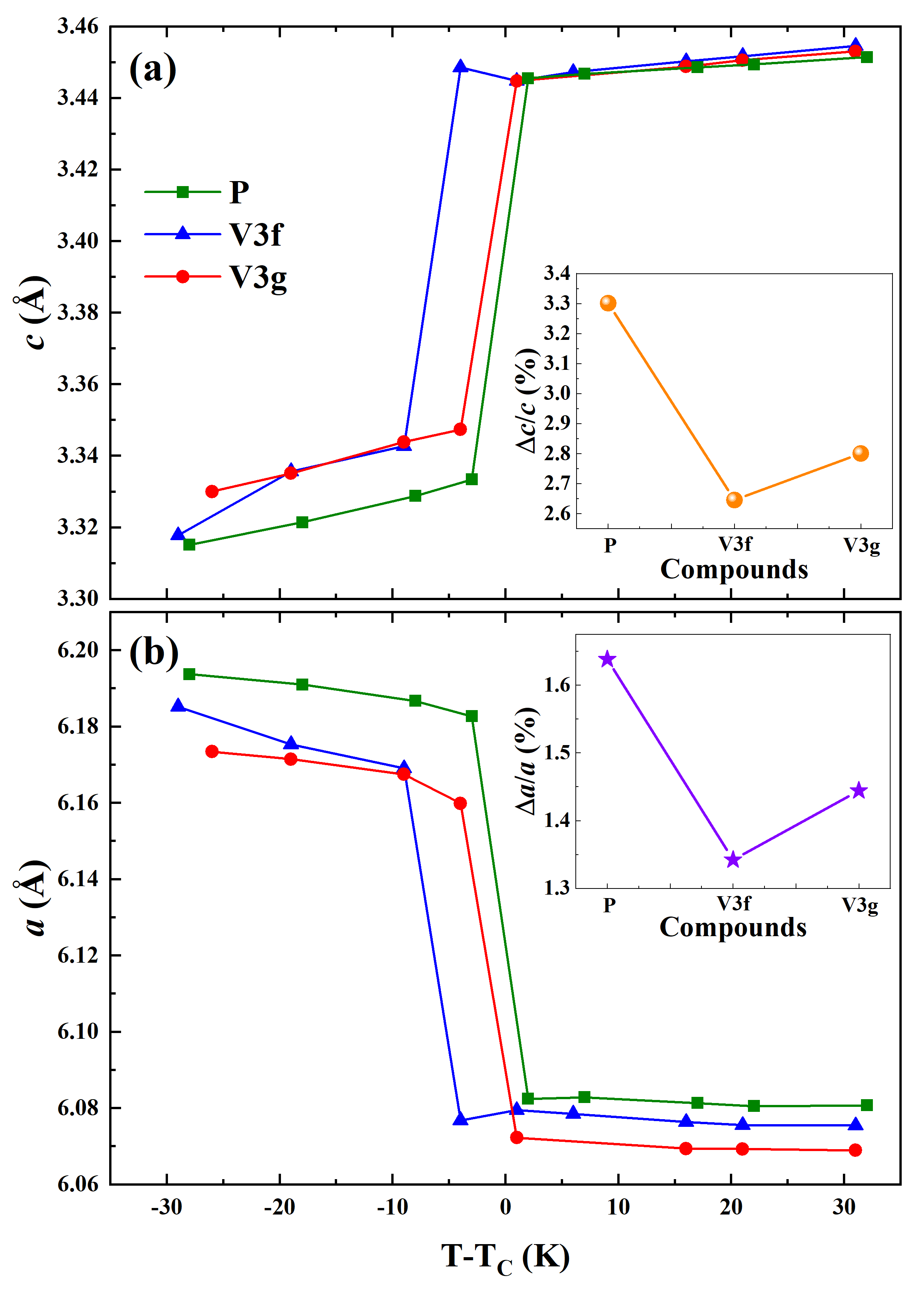}
    \caption{Temperature dependent variation of the hexagonal lattice parameters; (a) \textit{c} and (b) \textit{a} considering only the dominant Fe$_2$P-phase. The inset shows the relative variation of the lattice parameters at $T_C$. See supplementary for details. The error in the lattice parameter data is in the order of $10^{-4}$ Å, therefore not included in the figure.} 
    \label{ca}
\end{figure}

For a material to be useful in room-temperature magnetic refrigeration, the first requirement is to have a magnetic phase transition temperature near room temperature. The parent compound (FeMnP$_{0.5}$Si$_{0.5}$) has a $T_{C}$ value of around $380$ K. With 5 at$\%$ V substitution in either of the metallic sites (we will discuss later that V prefers the $3g$ site), $T_{C}$ decreases and comes closer ($\sim 320$ K) to room temperature (cf. Fig\ref{MCE}(c)), making this substitution process useful for magnetic refrigeration. The value of $T_{C}$ corresponds to the amount of thermal energy required to transform a material from its magnetically ordered state to a magnetically disordered state, therefore the value of $T_{C}$ is directly related with the strength of the exchange interaction between the spins of the magnetic atoms. Moreover, the exchange interaction between spins is highly sensitive to the inter spin distance. From neutron diffraction \cite{miao2014tuning,miao2016short} and Mössbauer spectroscopy \cite{miao2016kinetic} results, it has been observed that in the (Fe,Mn)$_{2}$(P,Si) system, Fe, Mn and P / Si occupy the $3f$, $3g$ and $1b$ / $2c$ sites of the hexagonal lattice, respectively. Among them, the magnetic atoms Fe and Mn are separated along the $c$-axis and distributed in the $ab$-plane. Hence, the distance along the $c$-axis represents the distance between Fe and Mn atoms. From temperature dependent XRPD results, the variation of the lattice parameter \textit{c} with temperature for every compound is depicted in Fig \ref{ca}(a). It should be kept in mind that near $T_{C}$ all compounds have two Fe$_2$P-type phases with different $c/a$ ratio. Among them the low $c/a$ ratio is dominant below $T_{C}$ and vice versa. Therefore, in the following discussion only the dominant Fe$_2$P-phase is considered. From Fig.\ref{ca}(a), it is observed that when the material transforms from PM to FM state, there is a large decrease (for the dominant  Fe$_2$P phase) of the lattice parameter $c$, indicating a decrease of the Fe to Mn atomic distance. From Table\ref{elasticTable} and inset of Fig.\ref{ca}(a), it is clear that the largest relative change of the $c$-parameter ($\Delta c/c$) has been observed for the parent compound, followed by the compounds V$3$g and V$3f$ in descending order. As expected, the $T_{C}$ values show the same trend. It can be concluded that the smaller the Fe to Mn distance along the \textit{c}-axis, the stronger the exchange coupling strength is and hence $T_{C}$ is inversely proportional to the Fe to Mn distance.

\begin{table}[ht]
\caption{Magnetic and magnetoelastic properties.}
\vspace{3pt}
\centering

\begin{ruledtabular}
\begin{tabular}{cccccc}
 Sample&   $T_{C}^{FC}$& $\Delta c/c$& $\Delta a/a$& $\Delta T_{hys}$&   $M_{S}$\\
   &(K)&(\%) &(\%)& (K)&($\mu_{B}/f.u.$)  \\\\
\hline
P&      376&   3.30&    1.64&    22&     4.04\\
V3f&    318&   2.64&    1.34&    8.5&     3.6\\
V3g&    322&   2.80&    1.44&    8.5&     3.7\\

\end{tabular}
\end{ruledtabular}

\label{elasticTable}
\end{table}

Apart from the variation of lattice parameter \textit{c}, the temperature dependent variation of the \textit{a} lattice parameter is shown in Fig.\ref{ca} (b). For the PM to FM phase transition a sharp increase of the \textit{a}-parameter is observed. Now, Fe and Si both occupy the same basal plane, thus an increase of the \textit{a}-parameter favors localization of the 3$d$ electrons and less bonding with Si atoms. From DFT calculation results \cite{dung2011mixed,boeije2016efficient} of (Fe,Mn)$_2$(P,Si)-system, it has been observed that the density of states (DOS) of Mn 3$d$ electrons remains identical in the FM and PM states while the DOS of the Fe 3$d$ electrons is significantly different comparing the FM and PM states. In fact the change of the local magnetic moment of Fe has been identified as the reason for observing the magnetoelastic coupling in the (Fe,Mn)$_2$(P,Si)-system \cite{miao2018overview}. Moreover, the increment of the \textit{a} lattice parameter, across the PM to FM transition represents a strong magnetoelastic coupling. Similar to the variation of the \textit{c}-parameter, the variation of \textit{a} ($\Delta a/a$, see Table\ref{elasticTable}) is also largest for the parent compound, followed by the V$3g$ and V$3f$ compounds. 
The large change of the lattice parameters \textit{a} and \textit{c} close to the magnetic transition is the reason for observing a strong first order magnetic phase transition for  these materials. The two Fe$_2$P-type phases, characterized by different \textit{a} and \textit{c} parameters are separated by an energy barrier, which is responsible for the temperature hysteresis ($\Delta T_{hys}$) of these materials \cite{dung2012magnetoelastic}. Table\ref{elasticTable} shows the variation of the lattice parameters ($\Delta a/a$ and $\Delta c/c$) across the magnetic transition indicating that the parent compound has the strongest magnetoelastic coupling or largest energy barrier, resulting in a relatively larger value of $\Delta T_{hys}$. On the contrary, the relatively lower values of $\Delta a/a$ and $\Delta c/c$ for the V$3f$ and V$3g$ compounds yield reduced values of $\Delta T_{hys}$. For the reversibility of a magnetic heat pump, a minimal value of $\Delta T_{hys}$ is required\cite{gutfleisch2016mastering}, indicating that V substitution constitutes a useful process for magnetic refrigeration applications.            

\subsection{Magnetization anomaly and M{\"o}ssbauer spectra}
Theoretical calculations of the site specific magnetic moment of the parent compound (FeMnP$_{0.5}$Si$_{0.5}$) show that the magnetic moments of Mn in the $3g$ site and Fe in the $3f$ site are $2.81~\mu_{B}/$atom and $1.68~\mu_{B}/$atom, respectively \cite{hudl2011strongly}. Therefore, substitution with nonmagnetic V is expected to reduce the overall magnetization more for $3g$-site substitution compared to $3f$-site substitution. However, an inverse behaviour is observed from the values of the saturation magnetization (cf. Fig\ref{MCE}(d) and Table\ref{elasticTable}). A possible reason of this anomaly could be a partly random occupancy of Fe and Mn atoms, i.e. if some amount of Fe (Mn) is distributed in the $3g$ ($3f$) site. To investigate this possibility, we have collected M{\"o}ssbauer spectra for the compounds in their paramagnetic states.

\begin{figure}[ht]
    \centering
    \includegraphics[width=\linewidth]{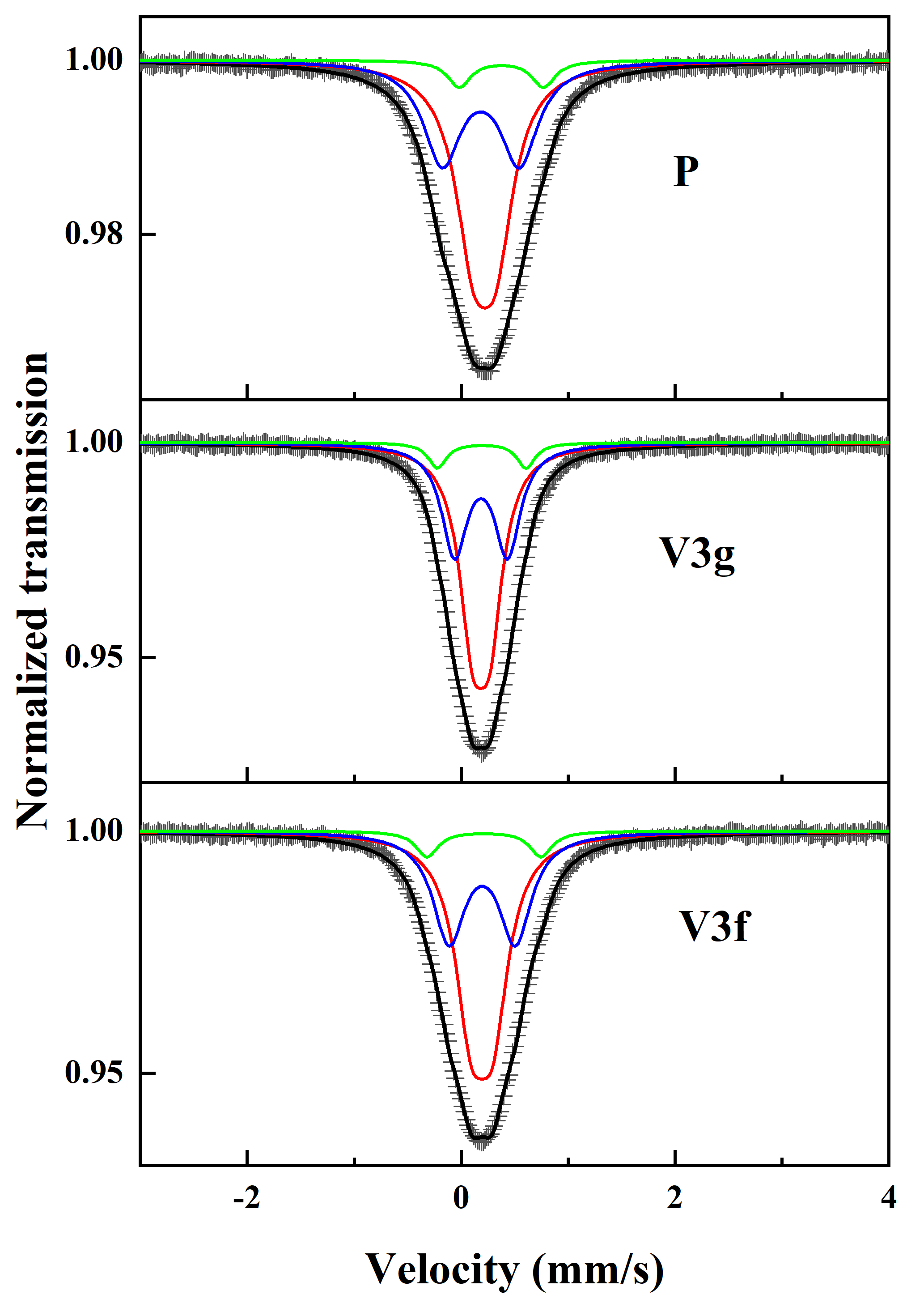}
    \caption{M{\"o}ssbauer spectra of the studied compounds at $410K$. The red, blue and green sub-patterns correspond to the nearest neighbour surroundings (P$_{2}$Si$_{2}$), (P$_{3}$Si$_{1}$) and (P$_{4}$) of Fe at the $3f$ site, respectively.} 
    \label{mos}
\end{figure}

In Fig.\ref{mos}, the M{\"o}ssbauer spectra of the three studied compounds are shown. For the parent compound, FeMnP$_{0.5}$Si$_{0.5}$, the broadenings emanate from the different surroundings of Fe at the metal $3f$ site.  There are four near neighbours elements P and Si, two occupying the $1b$ and two the $2c$ sites. It has been shown that Si prefers the $2c$ site almost exclusively \cite{fruchart2019structure}. For the present compound, two P atoms will occupy the two nearest $1b$ sites and assuming random occupation on the $2c$ sites we would expect three different near neighbour surroundings; P$_2$Si$_2$ (i.e. one Fe atom is surrounded by two P and two Si atoms), P$_3$Si$_1$  and P$_4$ with probabilities of $0.5625$, $0.375$, and $0.0625$, respectively. These components are shown in Fig.\ref{mos} with red, green and blue sub-patterns, respectively. Accordingly, the spectra at $410$ K, irrespective of V content were fitted with three doublets. The fitting results for the average hyperfine values are presented in Table\ref{mos_table}. The \textit{CS} values for the V substituted samples have decreased as compared to the value for the parent sample. It should be noted that a decrease in \textit{CS} corresponds to an increase in electron density at the Fe nuclei. This decrease in \textit{CS} can therefore be associated with the shrinking of the a-axis for the V substituted samples making the P and Si atoms in the first coordination sphere coming closer to the Fe nuclei. As discussed before, a higher value of the \textit{a}-parameter prevents Fe bonding with non-metallic atoms and yields the desired moment fluctuation across the PM to FM phase transition. A larger moment fluctuation will result in a larger change of magnetic entropy and hence a larger value of $-\Delta S_M$. The same trend for the values of \textit{CS} and $-\Delta S_M$ confirms the theoretical prediction \cite{miao2018overview} of the moment fluctuation of Fe in the $3f$ site. 

\begin{table}
\caption{Results from fitting of Mössbauer spectra.}
\vspace{3pt}
\centering

\begin{ruledtabular}
\begin{tabular}{cccc}
 Sample& $CS$ ($\pm 0.005$)& $\mid$\textit{QS}$\mid$ ($\pm 0.005$)& \textit{W} ($\pm 0.005$)\\
\hline
P&	0.220&	0.210&	0.406\\
V3g&	0.193&	0.336&	0.558\\
V3f&	0.185&	0.279&	0.443\\

\end{tabular}
\end{ruledtabular}

\label{mos_table}
\end{table}

The broad single line centered around $0.2$ mm/s matches well with results from a previous study \cite{hudl2011strongly} and evidences that Fe atoms occupy the $3f$ site. Also, the absence of any high velocity resonance line or shoulder diminishes the possibility of Fe $3g$ site occupation.           

\subsection{Chemical composition and magnetic phases}
The chemical compositions of the compounds as obtained from analysis of the EDX results are listed in Table\ref{EDXtable}.

\begin{figure*}[ht]
    \centering
    \includegraphics[width=\linewidth]{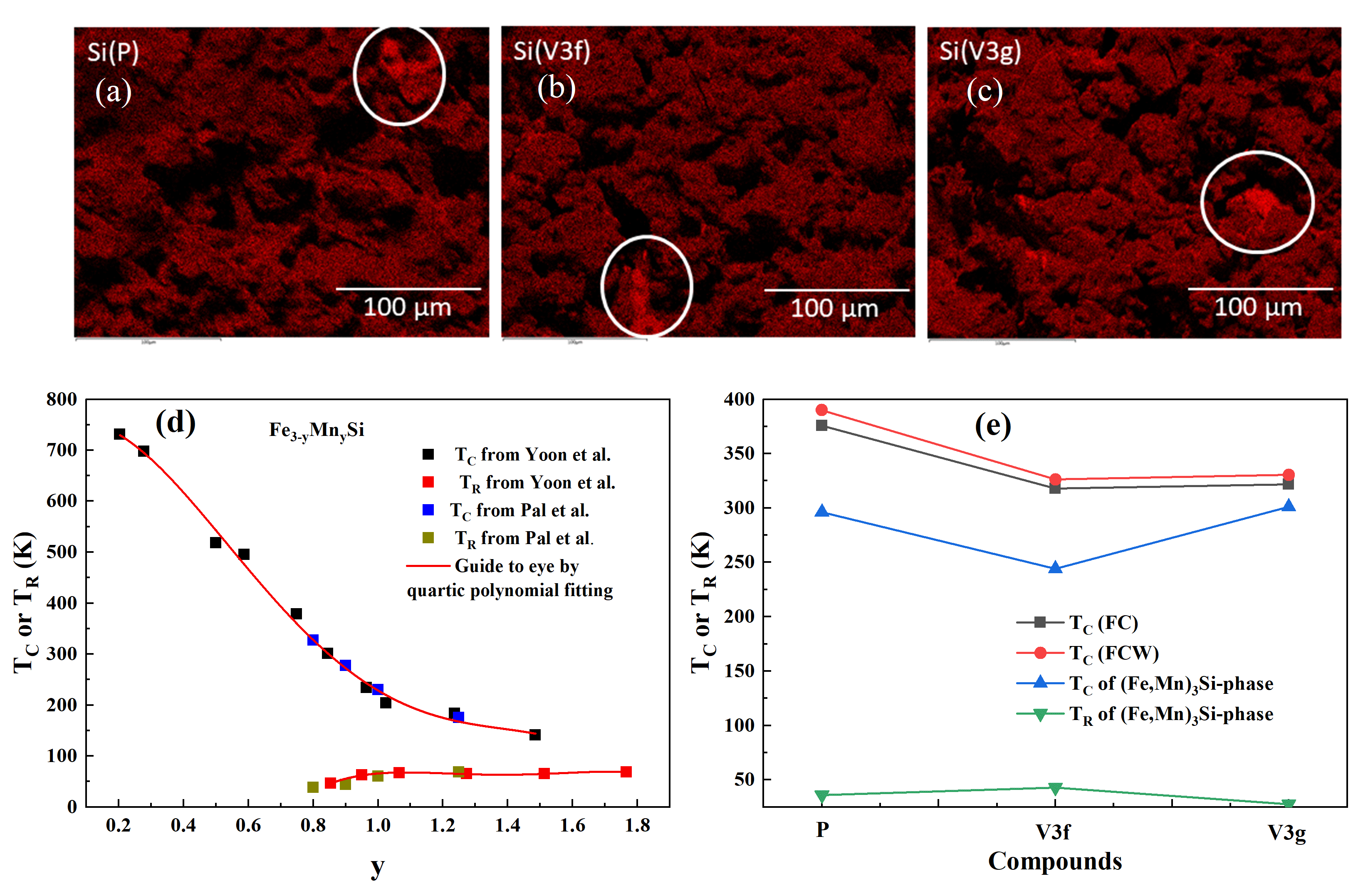}
    \caption{(a) - (c) Elemental mapping of Si for the three studied compounds. The circled regions indicate regions with excess of Si. (d) Magnetic phase diagram of   Fe$_{3-y}$Mn$_y$Si based on published literature values for the magnetic ordering temperatures. (e) $T_C$ and $T_R$ values for the primary and secondary phases of the studied compounds.} 
    \label{edx_impurity}
\end{figure*}

\begin{table}[ht]
\caption{Chemical composition of the studied compounds from EDX analysis.}
\vspace{3pt}
\centering

\begin{ruledtabular}
\begin{tabular}{cccc}
 Element (at\%)&    P&  V$3f$&  V$3g$\\
\hline
 Fe (expected)&    33.33&  31.67&  33.33\\
 Fe (observed)&    33(3)&  31(2)&  31(2)\\
 Mn (expected)&    33.33&  33.33&  31.67\\
 Mn (observed)&    35(3)&  34(2)&  31(1)\\
 V (expected)&     0&   1.67&  1.67\\
 V (observed)&     0&  1.9(6)&  1.7(2)\\
 P (expected)&    16.67&  16.67&  16.67\\
 P (observed)&    16(3)&  16(3)&  18(2)\\
 Si (expected)&    16.67&  16.67&  16.67\\
 Si (observed)&    15(2)&  17(2)&  19(2)\\

\end{tabular}
\end{ruledtabular}

\label{EDXtable}
\end{table}

From Table\ref{EDXtable} it is clear that all the compounds have the expected chemical composition within the margin of error. However, during analysis of the EDX results some Si-rich portions have been identified. As indicated in the supplementary section, from the elemental mapping of P it is clear that the above mentioned Si-rich portions exhibit P-deficiency. Typically, these Si-rich or P-deficient portions (cf. Fig.\ref{edx_impurity} (a)-(c)) correspond to a (Fe,Mn)$_3$Si phase. From the room temperature XRPD analysis a small amount ($\sim 5$ at\%) of (Fe,Mn)$_3$Si phase has been identified for the three compounds. Formation of this secondary phase indicates a possible loss of P during synthesis. However, the analysis of the EDX results exhibits a large error bar (as high as $3$ at$\%$ for the Fe and Mn content) and the XRPD refinement with multiple phases is not very sensitive to Fe/Mn intermixing. Fortunately, the magnetic properties of the  secondary phase can be used to predict the Fe to Mn ratio. The (Fe,Mn)$_3$Si-type phase exhibits a transition to a ferromagnetic state at high temperature along with a low temperature ($<50$ K) antiferromagnetic type spin-reorientation temperature ($T_R$). Without Mn, the Fe$_3$Si phase has a $T_C$ value of around $800$ K and with Mn insertion $T_C$ rapidly decreases to values below room temperature \cite{yoon1974structural,pal2013effect}. A magnetic phase diagram using literature values of the (Fe,Mn)$_3$Si phase is shown in Fig.\ref{edx_impurity}(d) and results of magnetic transition temperatures for the primary and secondary phases of our studied compounds as obtained from temperature dependent magnetization measurements are shown in Fig.\ref{edx_impurity} (e). Comparing the measured transition temperatures of the (Fe,Mn)$_3$Si phase with the transition temperatures shown in the phase diagram, it can be concluded that the secondary phase of the V$3f$ compound has a higher Mn to Fe ratio compared to the parent and V$3g$ compounds. This also indicates that the V$3f$ compound has a deficiency of Mn in the primary Fe$_2$P-type phase. Interestingly, all three compounds have been synthesized using identical conditions. Therefore the loss of Mn in the V$3f$ compound should have some intrinsic origin.

\subsection{Phase formation energy}

To find the reason of Mn loss in the V$3f$ compound and to estimate the effect of V substitution, the total energies of the systems have been calculated using density functional theory. For a more stable compound, the formation energy is expected to be negative and smaller relative to the pure components in their ground state structures. For the calculation two cases have been considered.\\
\textbf{Case $1:$} All Fe (Mn) atoms occupy $3f$ ($3g$) sites. Therefore, the formation energy for $3f$ site substitution with \textit{x} amount of V can be represented as,

\begin{equation*}
\begin{split}
    \Delta F_{Fe_{1-x}V_{x}MnP_{0.5}Si_{0.5}}= E_{Fe_{1-x}V_{x}MnP_{0.5}Si_{0.5}}\\-(1-x)E_{Fe}-xE_V-E_{Mn}-0.5E_P-0.5E_{Si} .
\end{split}    
\end{equation*}

Similarly, the formation energy for $3g$ site substitution will be,

\begin{equation*}
\begin{split}
    \Delta F_{FeMn_{1-x}V_{x}P_{0.5}Si_{0.5}}= E_{FeMn_{1-x}V_{x}P_{0.5}Si_{0.5}}\\-E_{Fe}-(1-x)E_{Mn}-xE_V-0.5E_P-0.5E_{Si} .
\end{split}    
\end{equation*}

The energy difference between $3f$ and $3g$ site substitution can therefore be expressed as, 
\begin{equation*}
\begin{split}
    \Delta F_1= E_{Fe_{1-x}V_{x}MnP_{0.5}Si_{0.5}}- E_{FeMn_{1-x}V_{x}P_{0.5}Si_{0.5}}\\+xE_{Fe}-xE_{Mn} .
\end{split}    
\end{equation*}

The value of $\Delta F_1$ for two different $c/a$ ratios are listed in Table\ref{tableEnergy}. Positive values indicate smaller formation energy for $3g$ site substitution. One may also note that the formation energy difference is higher for a higher level of V substitution. All these facts show that V prefers to occupy the $3g$ site.\\
\textbf{Case $2:$} Although it is known from analysis of the M{\"o}ssbauer results that Fe prefers the occupy the $3f$ site, we have no direct evidence that Mn can not occupy the $3f$ site. In this particular case a random occupancy of Fe and Mn in the $3f$ and $3g$ sites with V substitution is considered. Moreover, equimolar amounts in the $3f$ and $3g$ sites are considered, i.e. for each $1$ mol in total of P and Si, the total amount of Fe, Mn and V in the metallic sites will be $2$ mol. Therefore, similar to case $1$, the formation energies for $3f$ and $3g$ site substitutions have been calculated and their difference can be expressed as,
\begin{equation*}
\begin{split}
    \Delta F_2= E_{(Fe_{1-x}V_{x})(Fe_{x/2}Mn_{1-x/2})P_{0.5}Si_{0.5}}\\-E_{(Fe_{1-x/2}Mn_{x/2})(Mn_{1-x}V_{x})P_{0.5}Si_{0.5}} .
\end{split}    
\end{equation*}

\begin{table}[ht]
 \caption{Results from formation energy calculations. The energy differences are given in units of mRy/atom}
\vspace{2pt}
 \centering
 \begin{ruledtabular}
 \begin{tabular}{ccccc}
    x  &   $\Delta F_1$& $\Delta F_1$& $\Delta F_2$& $\Delta F_2$ \\
    & $(c/a=0.53)$&  $(c/a=0.58)$&  $(c/a=0.53)$&  $(c/a=0.58)$ \\ 
    \hline
          0  &  0.000&  0.000&  0.000&  0.000\\
       0.01  &  0.947&  0.681&  0.949&  0.969\\
       0.02  &  1.797&  1.328&  1.849&  1.391\\
       0.03  &  2.630& 1.946&   2.735&  2.079\\
       0.04  &  3.367& 2.546&   3.511&  2.759\\
       0.05  &  4.182& 3.126&   4.347&  3.373\\
 \end{tabular}
 \end{ruledtabular}
    \label{tableEnergy}
\end{table}

\begin{figure}[H]
    \centering
    \includegraphics[width=\linewidth]{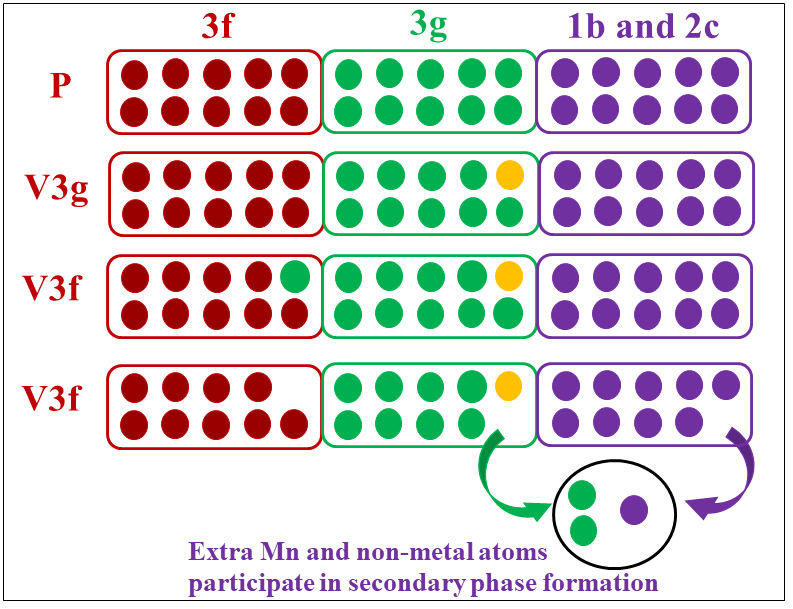}
    \caption{Model of V (yellow circles) substitution in the Fe (red circles) and Mn (green circles) dominated $3f$ and $3g$ sites, respectively. The purple circles represent P and Si. For the V$3$f compound two cases with Mn occupying the $3f$ site and Mn leaving the Fe$_2$P-phase are shown.}
    \label{model}
\end{figure}

Table\ref{tableEnergy} lists the values of $\Delta F_2$ for two different $c/a$ ratios. Similar to case $1$, case $2$ also indicates that V prefers to occupy the $3g$ site instead of the $3f$ site. To understand the physical consequences of this, a simplified model (cf. Fig.\ref{model}) is considered. $10$ atoms each of Fe and Mn are considered to occupy the $3f$ and $3g$ sites, respectively as a ground state (i.e. parent compound). Now, if $1$ Mn atom is replaced by $1$ V atom (i.e. V$3$g compound), following the total energy minimum criterion, V will occupy the $3g$ state. As a result there will be equimolar amount of Fe and Mn+V in the $3f$ and $3g$ sites, respectively. However, for the V$3f$ compound, the V atom will not occupy the $3f$ site, it will occupy the $3g$
site. This can have two consequences, either one Mn atom can occupy the $3f$ site or an equimolar amount of metallic atoms will occupy the $3f$ and $3g$ sites and the extra Mn atom will leave the Fe$_2$P phase and contribute to the secondary phase formation, as indicated in Fig.\ref{model}. In the first scenario, the Mn in the $3f$ site will interact antiferromagnetically with the Fe in $3f$ site \cite{delczeg2014origin}. In the second scenario, some amount of Mn will leave the Fe$_2$P phase of the V$3$f compound, and participate in the secondary phase formation, which will enhance the Mn/Fe ratio of the secondary phase (Fe$_{1-x}$Mn$_x$Si). The enhancement of the Mn/Fe ratio in the V$3f$ compound has been discussed previously. Moreover, in both cases, the overall magnetization of the V$3f$ compound will decrease, which explains the observed magnetization anomaly in the saturation magnetization result.

\section{Summary and conclusions}

V substitution in the metallic sites of FeMnP$_{0.5}$Si$_{0.5}$, results in a decrease of $T_C$, which is proportional to the magnetic exchange coupling strength. The above mentioned coupling strength is inversely proportional to the Fe to Mn distance along the hexagonal \textit{c} axis and proportional to the Fe to non-metal (P/Si) distance along the hexagonal \textit{a} axis. From the formation energy calculations, it was found that $3g$ site substitution is energetically favourable for the V atom. Attempting a $3f$ site substitution  will provoke either antiferromagnetic interaction in the $3f$-site or a secondary phase formation with the cost of an overall decrease of the magnetization. 
From M{\"o}ssbauer spectroscopy of the studied compounds apart from the absence of Fe in the $3g$ site, a decrease of the hyperfine parameter \textit{CS} (central shift) with V substitution has been observed. A larger \textit{CS} parameter represents non-bonded or weekly bonded Fe, which is favourable for the Fe moment fluctuation across the PM-FM phase transition \cite{miao2018overview}. A larger moment fluctuation results in a larger value of $-\Delta S_{M}$. Therefore, a direct correlation beween the \textit{CS} parameter and the value of $-\Delta S_{M}$ has been evidenced. Interestingly, the value of the temperature hysteresis $\Delta T_{hys}$ decreases with V substitution. The $\Delta T_{hys}$ in the Fe$_2$P-type systems originates from the energy barrier between the states  characterized by different $c$ and $a$ lattice parameters (described by $\Delta a/a$ and $\Delta c/c$). Here, in this work we have shown that with V substitution the energy barrier decreases considerably and results in a decrease of $\Delta T_{hys}$, which is highly desirable for the magnetic refrigeration application.\\    

\begin{acknowledgments}
The authors thank the Swedish Foundation for Strategic Research (SSF), project "Magnetic materials for green energy technology" (contract EM-16-0039) for financial support. Financial support from the Swedish Research Council (VR, contract 2019-00645) is gratefully acknowledged. The authors acknowledge support from STandUPP and eSSENCE. The computational studies were performed on resources provided by the Swedish National Infrastructure for Computing (SNIC).

\end{acknowledgments}

\bibliography{main.bib}
\bibliographystyle{main.bst}

\begin{widetext}

\clearpage

\appendix

\section{Supplementary Information}

\subsection{$\Delta c$, $\Delta a$, and $\Delta c/a$ calculation}
Fig.\ref{c_calculation} shows the \textit{c} lattice parameter values for the dominant ($>80\%$) Fe$_2$P phase above and below $T_C$ for the parent compound together with linear fits to describe the temperature dependence. The separation between the  fitted lines at $T_C$, represents the $\Delta c$ parameter. Similarly, the values of $\Delta a$ and $\Delta(c/a)$ have been calculated for the three compounds.

\begin{figure}[ht]
    \centering
    \includegraphics[width=0.7\linewidth]{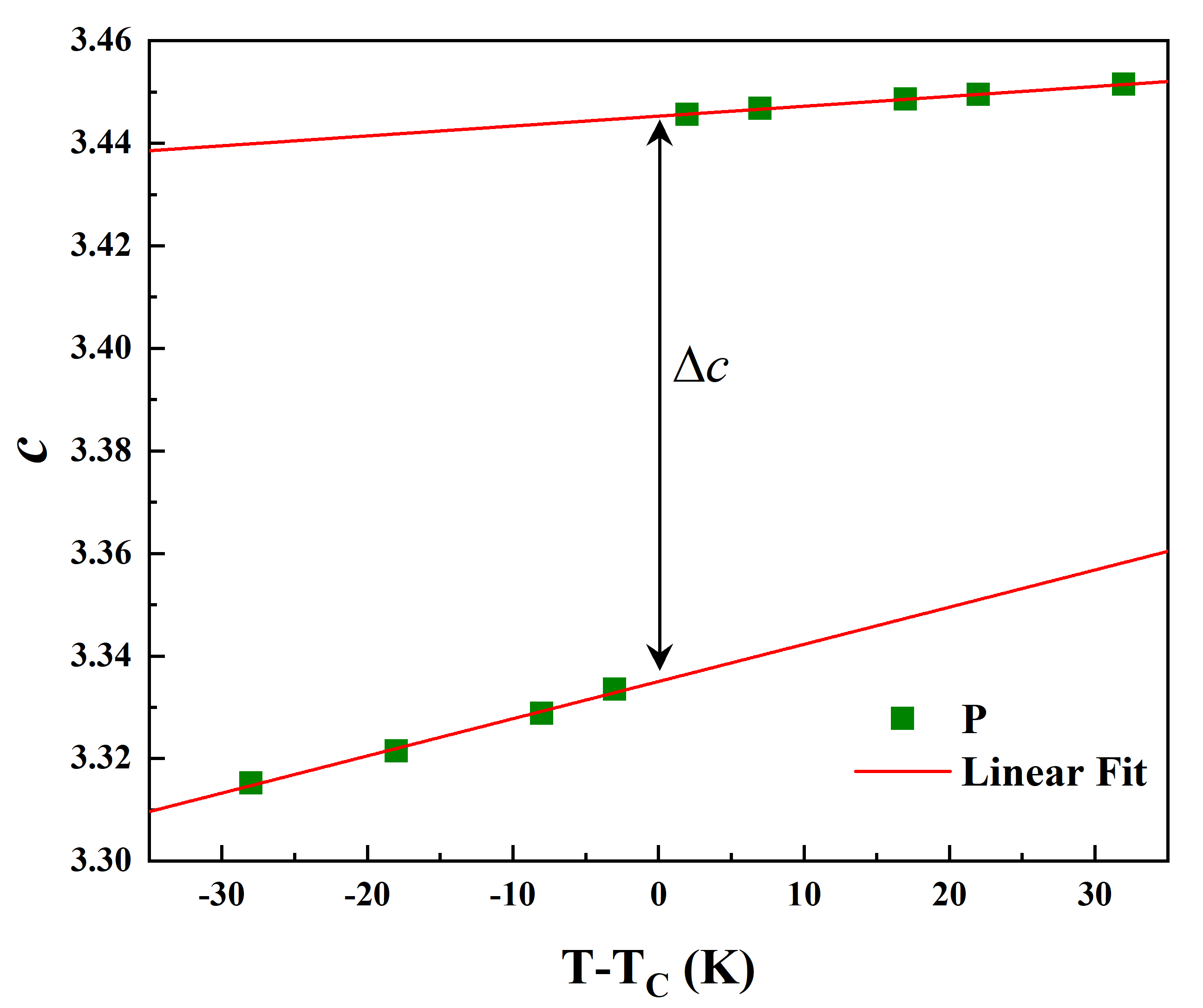}
    \caption{Temperature dependence of the \textit{c} lattice parameter for the dominant  Fe$_2$P phase for the parent compound showing how $\Delta c$ has been defined. Similarly $\Delta c$, $\Delta a$ and $\Delta(c/a)$ have been calculated for all the compounds.} 
    \label{c_calculation}
\end{figure}

\subsection{Temperature dependent XRPD}

\begin{figure}[H]
    \centering
    \includegraphics[width=\linewidth]{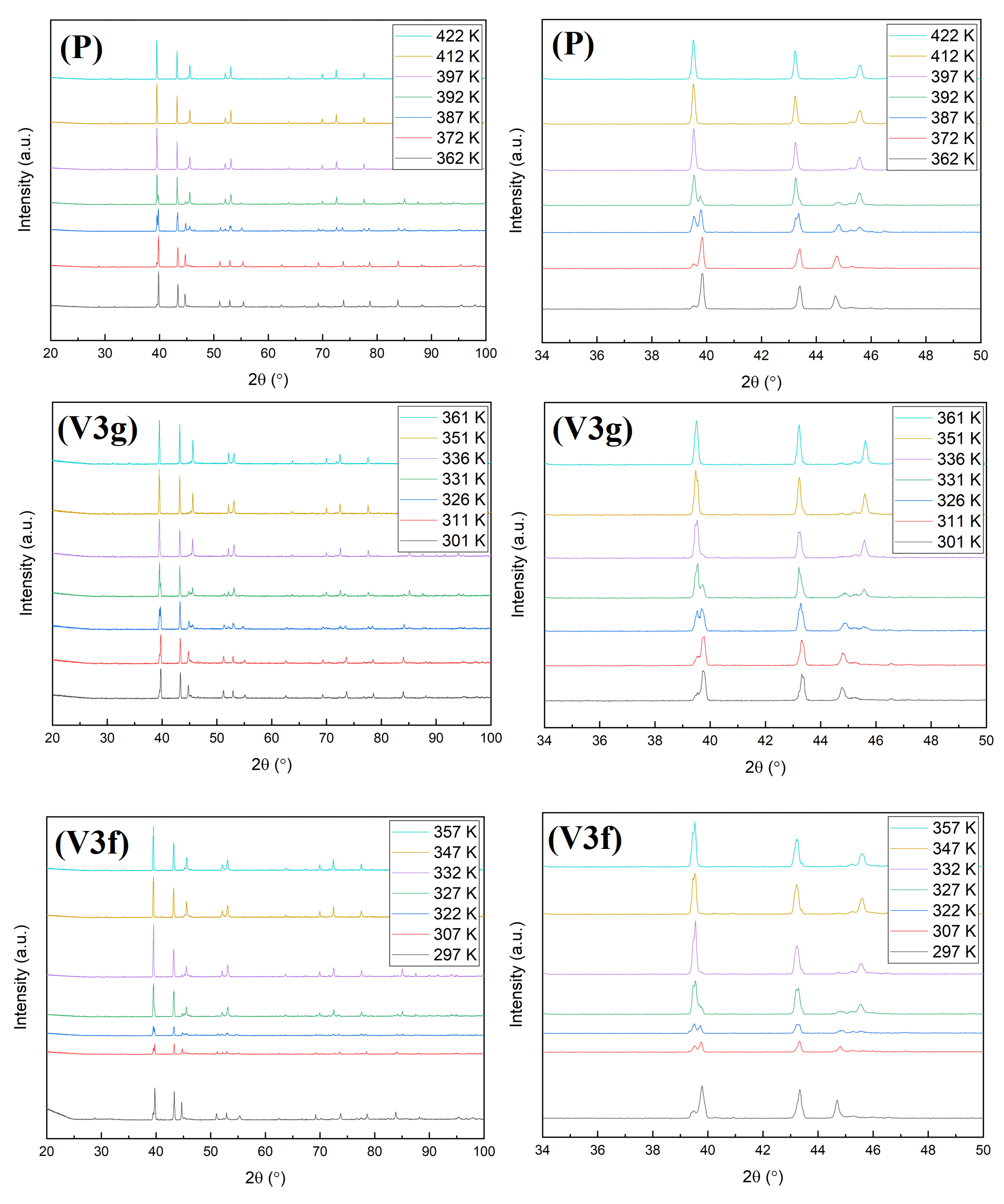}
    \caption{Temperature dependent XRPD patterns for the three studied compounds.} 
    \label{XRD_all}
\end{figure}

\subsection{ EDX mapping}

\begin{figure}[H]
    \centering
    \includegraphics[width=\linewidth]{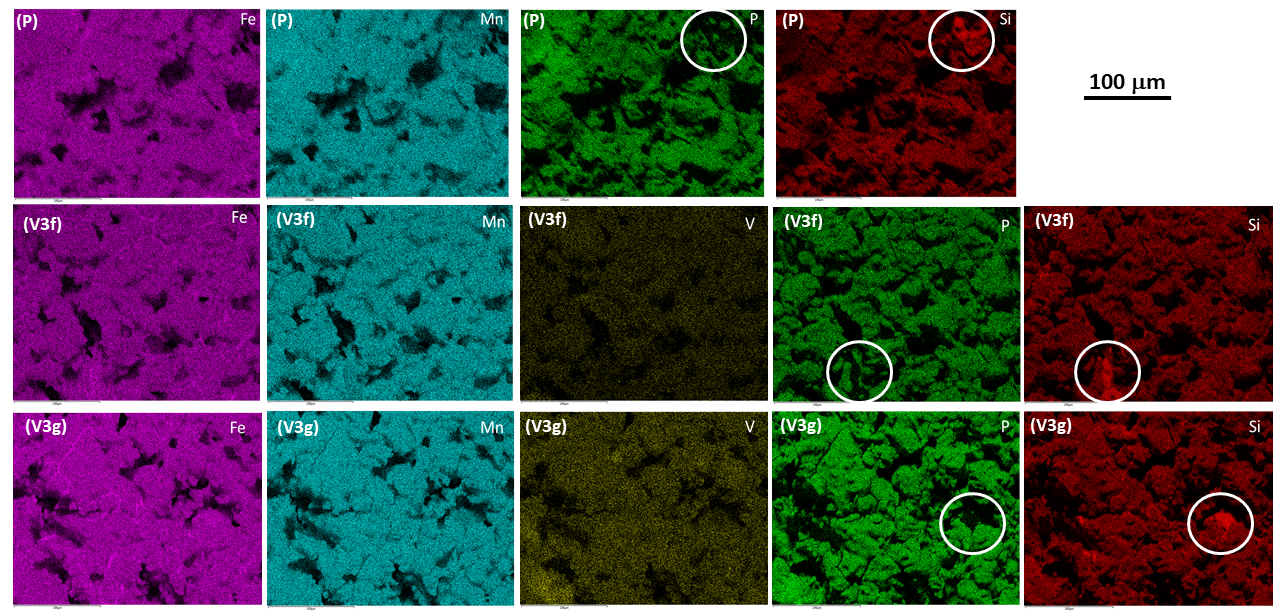}
    \caption{Elemental mapping of the studied compounds. The circled regions show deficiency of P and excess of Si.} 
    \label{edx_map}
\end{figure}
\end{widetext}

\end{document}